\documentclass[lettersize,journal]{IEEEtran}
\usepackage{amsmath,amsfonts}

\usepackage{algorithm}
\usepackage{algpseudocode}
\usepackage{array}
\usepackage[caption=false,font=normalsize,labelfont=sf,textfont=sf]{subfig}
\usepackage{textcomp}
\usepackage{stfloats}
\usepackage{url}
\usepackage{verbatim}
\usepackage{graphicx}
\usepackage{cite}
\usepackage{makecell}
\usepackage{multirow}
\usepackage{listings}
\usepackage{xcolor}
\usepackage{makecell}
\usepackage{booktabs} 
\usepackage{orcidlink}
\lstdefinelanguage{MyLang}{
    keywords={couple, part, parameter, port, value, connection,state,discrete,equation}, % 主要关键词
    emph={PartName1,PartName2, ParameterName1, CouplePortName1, VariableName1, DiscretePortName1 , InitialStateName , StateAction, StateName1 }, % 额外关键词
    keywordstyle=\color{blue}, % 主要关键词颜色
    emphstyle=\color{orange}, % 额外关键词颜色
    sensitive=false, % 是否区分大小写
}

\begin{document}

\title{GenAI for Simulation Model in Model-Based Systems Engineering}
\author{
Lin Zhang\orcidlink{0000-0003-1989-6102}\textsuperscript{*},~\IEEEmembership{Senior Member,~IEEE}\thanks{Lin Zhang, Yuteng Zhang, Pengfei Gu, and Zhen Chen are with School of Automation Science and Electrical Engineering, Beihang University, Xueyuan Road No.37, Haidian, Beijing, China; Hangzhou International Innovation Institute, Beihang University, 166 Shuanghongqiao Street, Pingyao Town, Yuhang District, Hangzhou, China; and State Key Laboratory of Intelligent Manufacturing Systems Technology, Yongding Road No.51, Haidian, Beijing, China. Lei Ren and Yuanjun Laili are with School of Automation Science and Electrical Engineering, Beihang University, Xueyuan Road No.37, Haidian, Beijing, China and State Key Laboratory of Intelligent Manufacturing Systems Technology, Yongding Road No.51, Haidian, Beijing, China. Dusit Niyato and Wentong Cai are with College of Computing and Data Science, Nanyang Technological University, Block N4-02a-32, Nanyang Avenue, Singapore. Agostino Bruzzone are with University of Genoa, Via Balbi, 5, 16126 Genoa, Italy.},
\and
Yuteng Zhang,
\and
Dusit Niyato\orcidlink{0000-0002-7442-7416},~\IEEEmembership{Fellow,~IEEE},
\and
Lei Ren\orcidlink{0000-0001-6346-6930},~\IEEEmembership{Senior Member,~IEEE},
\and
Pengfei Gu,
\and
Zhen Chen,~\IEEEmembership{Graduate Student Member,~IEEE},
\and

Yuanjun Laili,~\IEEEmembership{Member,~IEEE},
\and
Wentong Cai,~\IEEEmembership{Senior Member,~IEEE},
\and
Agostino Bruzzone
\thanks{This work was supported by Beijing Natural Science Foundation - Fengtai Innovation Joint Fund (Grant No. L241018) and Beihang World TOP University Cooperation Program. 
The authors would like to express their gratitude for the financial support provided by the funds.}
\thanks{This work has been submitted to the IEEE for possible publication. Copyright may be transferred without notice, after which this version may no longer be accessible.}
}

\markboth{Journal of \LaTeX\ Class Files,~Vol.~14, No.~8, August~2021}%
{Shell \MakeLowercase{\textit{et al.}}: A Sample Article Using IEEEtran.cls for IEEE Journals}


\maketitle

\begin{abstract}
Generative AI (GenAI) has demonstrated remarkable capabilities in code generation, and its integration into complex product modeling and simulation code generation can significantly enhance the efficiency of the system design phase in Model-Based Systems Engineering (MBSE). In this study, we introduce a generative system design methodology framework for MBSE, offering a practical approach for the intelligent generation of simulation models for system physical properties. First, we employ inference techniques, generative models, and integrated modeling and simulation languages to construct simulation models for system physical properties based on product design documents. Subsequently, we fine-tune the language model used for simulation model generation on an existing library of simulation models and additional datasets generated through generative modeling. Finally, we introduce evaluation metrics for the generated simulation models for system physical properties. Our proposed approach to simulation model generation presents the innovative concept of scalable templates for simulation models. Using these templates, GenAI generates simulation models for system physical properties through code completion. The experimental results demonstrate that, for mainstream open-source Transformer-based models, the quality of the simulation model is significantly improved using the simulation model generation method proposed in this paper. 
\end{abstract}

\begin{IEEEkeywords}
Generative AI, modeling \& simulation, model-based systems engineering
\end{IEEEkeywords}

\section{Introduction} \label{sec1}
\IEEEPARstart{T}{he} increasing demand for research and development of complex products, such as spacecraft and complex mechanical equipment, has placed greater demands on system design and verification. Model-Based Systems Engineering (MBSE) \cite{bib35} offers effective guidance for managing the complexity of such product development. The core principle of MBSE is to support all phases of a system's lifecycle—from the initial design stage, through the operation and maintenance stage, to the final disposal stage—by a unified, formalized, and standardized model, thus enabling the transition from a document-based research and development (R\&D) approach to a model-driven R\&D approach.

The three core elements of MBSE are modeling languages, tools, and methodologies. Graphical system modeling languages, such as the Systems Modeling Language (SysML) \cite{bib10}, are widely used in academia and industry to support system engineering views and define various R\&D elements in complex product development. However, mainstream MBSE methods, such as Harmony-SE \cite{bib11} and Magicgrid \cite{bib12}, face challenges when using various graphical modeling languages, particularly in addressing system architectures and physical properties. The different characterization methods of system architectures and physical properties lead to poor model consistency. Although metamodel transformation \cite{bib55} and co-simulation \cite{bib54} provide promising solutions, both approaches require the collaborative use of multiple heterogeneous languages and tools. As a result, the current MBSE-oriented R\&D approach increases the learning curve and inefficiency for system engineers. Furthermore, this multi-language and tool-integrated development approach restricts the application of intelligent methods to automate the generation of multi-level models throughout the system development process.

Unifying the modeling and simulation process is a key trend in the evolution of MBSE. A unified model representation can effectively address the challenges of integrating multi-stage and multi-domain models, thereby enhancing the efficiency of complex product research and development. Numerous institutions and scholars have explored integrated modeling and simulation languages, tools, and methodologies. For example, organizations like the International Council on Systems Engineering are advancing the adoption of the semantic modeling language SysML 2.0 \cite{bib46}, which aims to replace the current SysML. This new version enhances model expressiveness and consistency through a unified meta-modeling framework. The KARMA specification, a multi-architecture unified modeling language along with its associated toolchain system, proposed by Lu et al. \cite{bib47}, has been utilized to address the data integration challenges during the development of complex equipment. In our previous research, we proposed an integrated modeling and simulation language based on discrete event system specification (DEVS) \cite{bib29} extension development— X language \cite{bib17,bib18}. By extracting the structural representations of mainstream modeling languages, X language identifies the commonalities and interaction mechanisms across multi-domain models, providing a unified characterization approach for system architecture and physical properties. Additionally, X language defines two modeling forms—graphical and textual—consistent with meta-models. The aforementioned features of X language provide a strong foundation for intelligent generative design and modeling. 

Generative artificial intelligence (GenAI) has enabled the integration of intelligence into system modeling and simulation within MBSE. GenAI, in its primary form as a large language model (LLM), is grounded in deep learning techniques and can perform various natural language processing tasks. With computing power advances, several ultra-large-scale models have emerged, such as GPT3 \cite{bib2} and LLama2 \cite{bib3}. With billions or even hundreds of billions of parameters, these models can handle larger data sizes and more complex tasks, and are widely used and integrated into software systems in several domains. AI models' maturity and adaptability in coding scenarios have also grown significantly. GPT-4 \cite{bib21} is an upgraded version of GPT-3. In addition to the basic complementary functions, GPT-4 is able to write code directly and help users debug, and its code has been greatly improved in both quality and length. Open-source code generation models in the same period include Code-Llama \cite{bib22} by meta, WizardCoder \cite{bib23} proposed by Microsoft, Code-Qwen \cite{bib24} model introduced by Alibaba Group and CodeGeeX \cite{bib25} proposed by Wisdom Spectrum AI, etc. The rapid development of LLM in terms of information extraction and code generation has made possible its application in industry \cite{bib30, bib57} and MBSE. The rich knowledge base and strong innovation capabilities of LLMs enable them to generate system code automatically, relationships between model elements, and related documentation, accelerating the system design and development process and increasing the productivity of engineers. Cámara, J. et al. \cite{bib7} investigated ChatGPT's current ability to perform modeling tasks and assist modelers, Tikayat Ray et al. \cite{bib9} used an LLM to standardize a requirements hierarchy model, and Bader, E. et al. \cite{bib40} implemented the task of generating UML component diagram elements by fine-tuning a large language model. However, most related GenAI applications focus primarily on the requirement and functional levels of MBSE, with limited research on the generation of simulation models for the system physical properties of products. The simulation model for the system physical properties of products represents one of the key models in MBSE. It is the foundation for achieving critical tasks such as life prediction, fault diagnosis, and performance evaluation.

Based on the previous research, this paper proposes an innovative generative system design methodology framework for MBSE, offering a practical direction for the intelligent generation of simulation models focused on the system physical properties of products. We employ inference techniques, generative models, and integrated modeling and simulation languages to construct simulation models for system physical properties based on product design documents. The scalable templates for simulation models proposed in this paper ensure the accuracy of the generated models. The evaluation framework proposed in this paper provides a practical approach for evaluating the quality of generated models. The main contributions of this paper are as follows:
1) This paper proposes a generative system design method for MBSE. The method utilizes BERT, Transformer-based models, and X language to extract information regarding the model's composition, architecture, and behavior from product design documents, subsequently constructing simulation models for system physical properties.
2) A specification and method for constructing scalable simulation model templates are proposed. Using these templates, Transformer-based models generate simulation models for system physical properties through code completion. This approach overcomes LLM's limitations in processing long text inputs. 
3) A simulation model evaluation method is proposed for simulation models generated by Transformer-based models. The method introduces evaluation metrics beyond code accuracy tailored to the unique characteristics of simulation models and employs the entropy weighting method (EWM) to calculate weights, enhancing the evaluation's objectivity.

The structure of this paper is organized as follows: Section~\ref{sec2} describes the construction specifications for X-language-based scalable templates and the method for constructing scalable templates at different model levels. Section~\ref{sec3} details three key methods: 1) the method using BERT, Transformer-based models, and scalable templates to generate simulation models for system physical properties based on the product design documents, and 2) the evaluation method for the generated simulation model. Section~\ref{sec4} outlines the complete process of generating and evaluating the aircraft electrical system simulation model using the aforementioned simulation model generation methods. 

The definitions of key terms involved in this paper are summarized in Table.~\ref{mytab11} to avoid ambiguity. 

\begin{table*}[ht]
\centering
\caption{Term and definition table} \label{mytab11}
\renewcommand{\arraystretch}{1.2} 
\begin{tabular}{|m{4.2cm}|m{13cm}|}
\hline
\textbf{Term} & \textbf{Definition} \\
\hline
simulation model & simulation model for system physical properties of products \\
\hline
header/attribute/connection & components of the couple class model \\
\hline
header/definition/state/equation & components of the atomic class model \\
\hline
Name/Import/Port/Part/Connection & keywords of X language couple class and the code governed by the keywords in X language couple class model \\
\hline
\makecell[l]{Name/Import/Port/Value/\\Parameter/State/Transform/Equation} & keywords of X language atomic class and the code governed by the keywords in X language atomic class model \\
\hline
correctness similarity & a metric for evaluating the degree of deviation of an incorrect simulation model from an ideally correct model \\
\hline
simulation correctness & the degree of correctness of the simulation model, calculated by correctness similarity \\
\hline
degree of error & a metric to quantify the actual impact of errors on the simulation model code \\
\hline
model consistency & a metric to evaluate the consistency of the same elements across upper and lower models. \\
\hline
scalable template & simulation model templates characterized by a flexible design pattern, allowing their length and content to be dynamically adjusted based on the number or attributes of the included elements \\
\hline
product design document & the natural language documentation that engineers use as a reference when designing and modeling a system \\
\hline
simulation model corpus & a corpus comprising content related to the design of simulation models \\
\hline
component-level model corpus & a corpus comprising content related to the design of component-level simulation models \\
\hline
\end{tabular}
\end{table*}

\section{X Language Simulation Model Template Construction} \label{sec2}

This section introduces the fundamental concepts of DEVS and presents X language, a formal modeling and simulation language based on DEVS. X language model will serve as the primary object generated by the method proposed in this paper. At the end of this section, we describe a scalable template of X language based on its syntax, which serves as a foundational prerequisite for the proposed method.

\subsection{Basic Concepts of DEVS}\label{subsec21}
DEVS represents a formal modeling specification used to describe discrete event systems. DEVS models typically comprise the following core components: 

\textbf{Atomic Model} defines the fundamental constituent units of a system, capturing its behavior and state. The atomic model in DEVS is defined by a seven-tuple constructor:
\begin{align}
{ AtomicDEVS }=<S, t a, \delta_{i n t}, X, \delta_{e x t}, Y, \lambda> \label{myeq21}
\end{align}

The elements of \eqref{myeq21} are defined as follows:
    1) State set (\(S\)): Describe all possible states of the model.
    2) Input Event Set (\(X\)): Define the types of external events the model can receive.
    3) Output Event Set (\(Y\)): Define the types of external events the model can generate.
    4) Internal Transition Function (\(\delta_{i n t}\)): Define the rules governing the transition of a model from one state to another, independent of external events.
    5) External Transition Function (\(\delta_{e x t}\)): Defines the procedure for updating the model's state upon receiving an external event.
    6) Output Function (\(\lambda\)): Defines the output behavior triggered by the model in a particular state.
    7) Time Advance Function (\(ta\)): Defines the duration for which the model stays in its current state before a state transition occurs.

\textbf{Coupled Model} integrates multiple atomic models or other coupled models to create more complex system structures.  The coupled model in DEVS is defined by a four-tuple constructor:
\begin{align}
{ CoupledDEVS }=<D, EIC, EOC, IC> \label{myeq22}
\end{align}

The elements of \eqref{myeq22} are defined as follows:
    1) Set of Components (\(D\)): Include multiple atomic models or other coupled models.
    2) External Input Coupling (\(EIC\)): Defines how external input events are passed to the submodel.
    3) External Output Coupling (\(EOC\)): Define how the output events of the submodel are passed to the outside of the system.
    4) Internal Coupling (\(IC\)): Define event transfer relationships between sub-models or between sub-models and external components.

The operational mechanism of the DEVS model is based on the event-driven principle and progresses in discrete time steps. Although the core design of the DEVS architecture primarily focuses on modeling discrete events, extensions of DEVS and some of its derivative frameworks support continuous-discrete hybrid systems by introducing mechanisms for ``continuous state change'' or by combining continuous state variables with discrete event processing  \cite{bib48}. Hybrid DEVS and similar extensions provide modeling capabilities for continuous-discrete hybrid systems, but current DEVS tools (e.g., CD++ and DEVSJAVA) offer limited support for hybrid modeling. Most of these tools remain focused on traditional discrete-event systems, and the hybrid modeling functionality is neither fully optimized nor standardized  \cite{bib49}. Some tools may necessitate users to develop additional modules for hybrid modeling or perform extensive custom configurations, thereby increasing development costs.

\subsection{Basic Concepts of X Language} \label{subsec22}

\begin{figure}[!t]
\centerline{\includegraphics[width=\columnwidth]{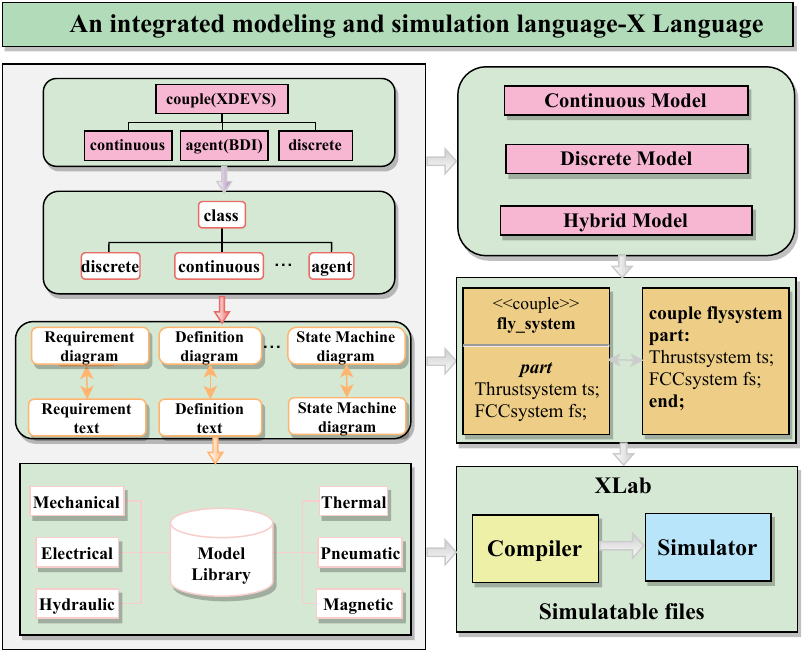}}
\caption{\raggedright Based on XDEVS (an extension of DEVS), X language defines various classes, including discrete, continuous, and agent classes, and supports both graphical and textual representations. X language supports modeling continuous, discrete, and hybrid models and enables the compilation and simulation of models using X language development tools, XLab.}\label{myfig17}
\end{figure}

The structure of the X language is illustrated in Fig.~\ref{myfig17}. X language is a modeling and simulation language developed based on DEVS. It complements the description of continuous port connectivity and integrates both continuous and discrete event port connectivity to model and simulate the interaction between continuous and discrete behavior-dominated hybrid models. X language supports the integrated modeling of system architecture and physical characteristics, thereby enabling comprehensive modeling and simulation analysis in system designs \cite{bib19,bib20}. Additionally, X language supports dual-mode modeling (graphical and textual), with graphical representation enhancing interactivity, whereas textualization enables the integration of GenAI within X language. 

The core classes of X language are presented in Table.~\ref{mytab23}. This section provides a detailed description of the functions of these classes.

\textbf{Couple class:} Couple class is a crucial component in X language, facilitating the integration of simulation analysis modeling and system architecture modeling. It corresponds to the coupled model in the DEVS architecture. The keyword Part in the attribute component and the keyword Connection in the connection component are specific to the couple class and used to describe the model's composition and the connection relationship, respectively, and together constitute the architecture of the couple class model.

\textbf{Discrete class:} Discrete classes facilitate the modeling of discrete-event systems. The discrete class is indivisible, serving as the fundamental simulation unit within a discrete model. The keywords State and Transform in the state component are specific to the discrete class and used to represent transfer scenarios and conditions between different states in discrete events.

\textbf{Continuous class:} Continuous class is established to accomplish the modeling of continuous systems. The continuous class facilitates the construction of non-causal models, primarily based on equation solving. The keyword Equation in the equation component is specific to the continuous class and used to describe the behavior of the continuous class.

\begin{table*}[ht]
\caption{Components of X language class}\label{mytab23}
\centering
\renewcommand{\arraystretch}{1.2} 
\begin{tabular}{|m{2cm}|m{2cm}|m{2cm}|m{10cm}|}
\hline
\textbf{Class} & \textbf{Component} & \textbf{keyword} & \textbf{definition} \\

\hline
\multirow{5}{*}{couple}&\multirow{2}{*}{header} & the Name & name of the couple class model \\ \cline{3-4}
                     &   & Import & the import of subsystem models (couple class or atomic class) \\ \cline{2-4}
&\multirow{2}{*}{attribute} & Part & the submodule contained in the couple, i.e., instantiation of the subsystem models \\ \cline{3-4}
                      &   & Port & the port of the couple class for connecting to external systems \\ \cline{2-4}
&connection            & Connection & the signal or data connection relationship between subsystems \\

\hline
\multirow{6}{*}{discrete}&\multirow{2}{*}{header} & Name & the name of the discrete class model \\ \cline{3-4}
                      &  & Import & the import of functions used in the model \\ \cline{2-4}
&\multirow{3}{*}{definition} & Port & the port of the discrete class for connecting to external systems \\ \cline{3-4}
                    &    & Value & the model's variables during the simulation process \\ \cline{3-4}
                    &     & Parameter & the intrinsic property of the discrete class where its internal elements are constants. \\ \cline{2-4}
&\multirow{2}{*}{state}    & State &  the transfer situation and transfer conditions between different states in a discrete event \\ \cline{3-4}
&& Transform & the transformation conditions and transformation processes of discrete states \\

\hline
\multirow{6}{*}{continuous} & \multirow{2}{*}{header} & Name & the name of the continuous class model \\ \cline{3-4}
                    &    & Import & the import of functions used in the model \\ \cline{2-4}
&\multirow{3}{*}{definition} & Port & the port of the continuous class for connecting to external systems\\ \cline{3-4}
    &                    & Value & the model's variables during the simulation process \\ \cline{3-4}
      &                   & Parameter & the intrinsic property of the continuous class where its internal elements are constants. \\ \cline{2-4}
&equation   & Equation & the dynamic behavior of the continuous class \\ \hline
\end{tabular}
\end{table*}

\begin{figure}[!t]
\centerline{\includegraphics[width=\columnwidth]{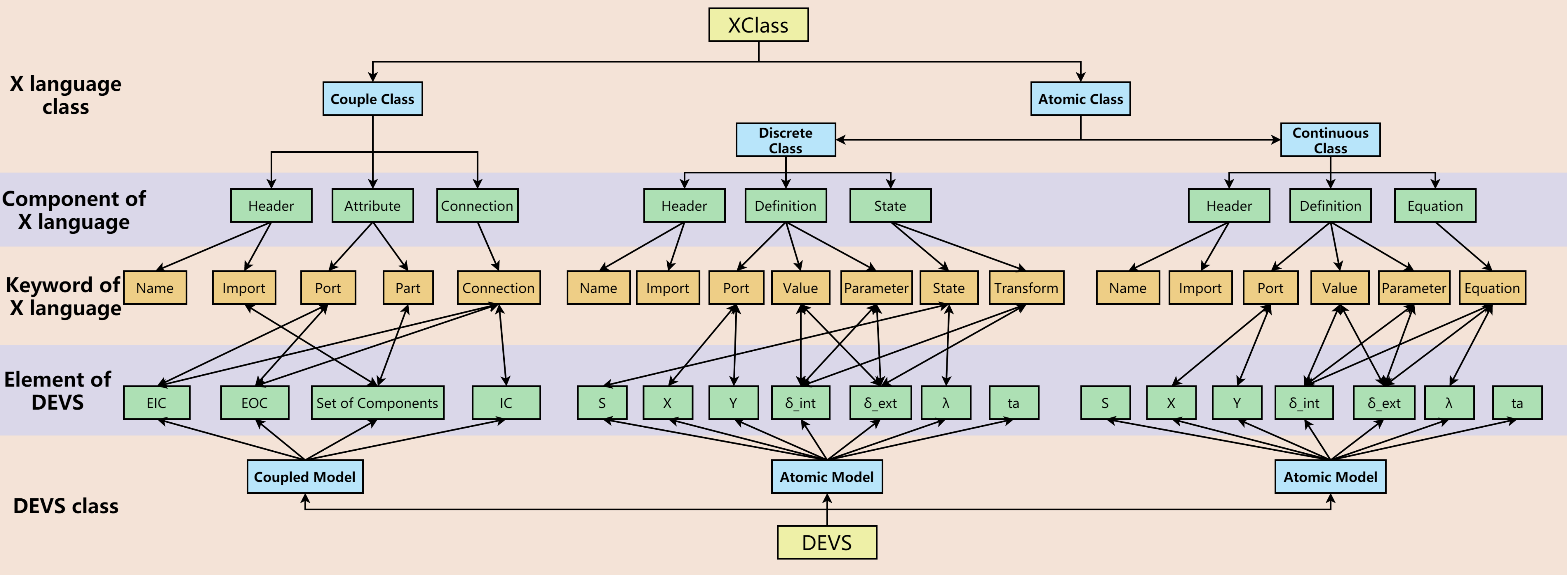}}
\caption{\raggedright Elements in X language classes can be mapped to corresponding elements in DEVS \cite{bib56}. Taking the coupled model of DEVS as an example, the functions of \(EIC\) and \(EOC\) are implemented through the keywords Port and Connection in the couple class of X language. The function of \(IC\) is implemented using the keyword Connection alone.}\label{myfig13}
\end{figure}

The continuous and discrete classes are atomic models in X language, corresponding to the atomic models in DEVS. The mapping between X language couple class, discrete class, and continuous class models and the corresponding elements in DEVS is illustrated in Fig.~\ref{myfig13}. X language couple class and atomic class models encompass all elements of the coupled and atomic models in DEVS, with some extensions beyond these elements.

\subsection{X Language Class Template Construction}\label{subsec23}

Complex product simulation models are characterized by high complexity and multi-scale. For such model code generation scenarios, generating accurate and consistent simulation code directly through GenAI is challenging. In this paper, the simulation model is constructed using scalable templates based on X language to address the issues above. X language scalable template deconstructs the syntax while preserving key elements and the overall model structure, thereby allowing model length and content to be dynamically adjusted based on the number and attributes of the included elements.

The core idea behind employing scalable templates for constructing simulation models is to modify the approach to code generation tasks. This study generates the simulation model through a modular code completion task. Compared to generating a complete simulation model, modular code completion can effectively address issues such as performance degradation, context loss, and output truncation caused by excessively long input contexts \cite{bib41}. However, modular code completion can reduce consistency from module to module and system to subsystem. Therefore, when generating simulation models using scalable templates, additional algorithms are required to enhance the consistency of the models, which will be explained in detail in Section~\ref{sec3}.

There are several rules to follow when building a scalable template:
    1) The top-down design principle should be followed when constructing the template. Specifically, model templates at the system level should be constructed prior to model templates at the component level.
    2) When constructing the template, the most effective description of the system-level structure of the simulation model and its contained elements should be extracted to minimize the amount of text generated by the LLM and improve accuracy.
    3) The coupling relationships between modules must be clarified when constructing the template. The coupling relationship of modules should be considered when using GenAI to generate simulation models.

Based on the characteristics of X language and the modeling specifications of the scalable templates discussed above, this paper develops scalable templates for X language couple class model, discrete class model, and continuous class model, as follows. A detailed description of the usage of these templates will be provided in Section~\ref{sec3}.

\begin{lstlisting}[]
couple <CoupleName>
  import <AtomicName1>;  
  import <AtomicName2>;  
  ...
part:   
  <AtomicName1> <PartName1>;  
  <AtomicName2> <PartName2>;
  ...
parameter:
  <DataType> <ParameterName1> = <Value1>;
  ...
port:
  <PortType> <CouplePortName1> = <InitialValue1>;
  ...
value:
  <DataType> <VariableName1> = <Value1>;
  ...
connection:
  connect(<PartName1>.<PortName1>, <PartName2>.<PortName2>);
end;

discrete <DiscreteName>
  import <FunctionName1>;
  ...
parameter:
  <DataType> <ParameterName1>=<Value1>;
  ...
value:
  <DataType> <VariableName1>=<Value1>;
  ...
port:
  <PortType> <DiscretePortName1>=<InitialValue1>;
  ...
state:
initial state <InitialStateName>
  when entry() then
    statehold(<StateholdTime>);
  end;
    <StateAction>
  end;

state <StateName1>
  when entry() then
    statehold(<StateholdTime>);
  end;
    <StateAction>
  end;
  ...
end;

continuous <ContinuousName>
  import <FunctionName>
  ...
parameter:
  <DataType> <ParameterName1>=<Value1>;
  ...
value:
  <DataType> <VariableName1>=<Value1>;
  ...
port:
  <PortType> <DiscretePortName1>=<InitialValue1>;
  ...
equation:
  <Equation>;
  ...
end;

\end{lstlisting}

\section{Simulation Model Generation Method Based on Scalable Templates and Transformer-based models} \label{sec3}

\begin{figure*}[!t]
\centerline{\includegraphics[width=0.65\textwidth]{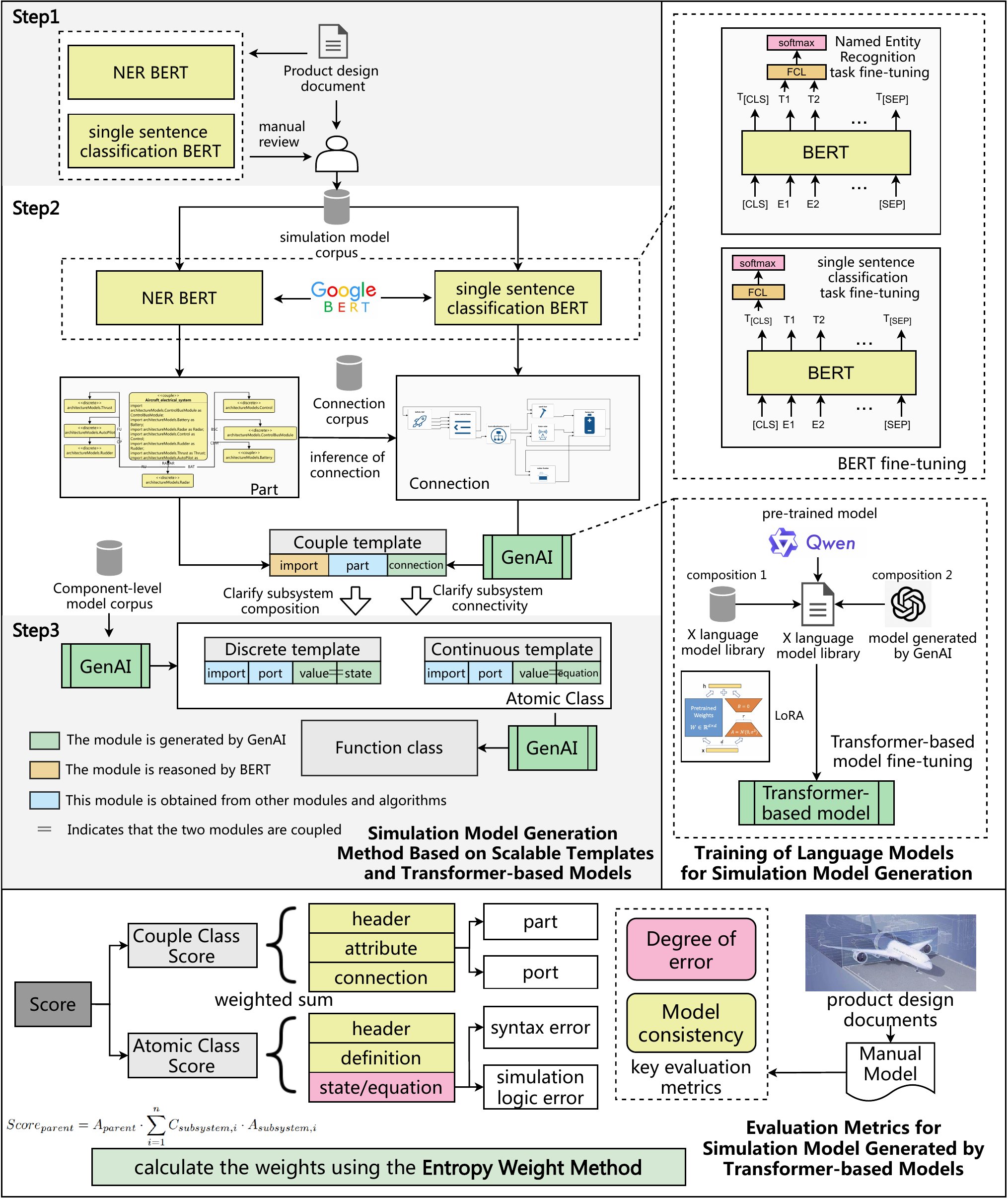}}
\caption{\raggedright The overall technical implementation framework is divided into three parts: Simulation Model Generation Method Based on Scalable Templates and Transformer-based Models, Training of Language Models for Simulation Model Generation, Evaluation Metrics for Simulation Models Generated by Transformer-based Models.}\label{myfig14}
\end{figure*}

This section focuses on integrating BERT, Transformer-based models, and scalable templates to generate corresponding simulation models from product design documents. The overall technical implementation framework for the proposed method is illustrated in Fig.~\ref{myfig14}. Specifically, Section~\ref{subsec31} provides a detailed introduction to the overall technical framework and specific steps of the simulation model generation process. Section~\ref{subsec32} and Section~\ref{subsec33} elaborate on the data processing and training methods of various language models employed in this workflow. Finally, Section~\ref{subsec34} proposes a set of evaluation metrics to evaluate the performance of different language models in generating simulation code.

\subsection{Overall Technical Implementation Framework}\label{subsec31}

This section elaborates on the three steps of the Simulation Model Generation Method Based on Scalable Templates and Transformer-based Models.

\textbf{Step 1: Identify the relevant corpus for target simulation models.}  Product design document refers to the natural language documentation that engineers use as a reference when designing and modeling a system. However, such documents may contain irrelevant information, such as background knowledge or excessive details, which are unnecessary for constructing a simulation model. Therefore, extracting the relevant information from these documents for model construction is essential. We use BERT fine-tuned for Named Entity Recognition (NER) and single-sentence classification tasks to process the documents. If a sentence is not labeled in either task, it is considered redundant and excluded from the corpus. The remaining sentences are then added to the simulation model corpus. Subsequently, the product design documents are manually reviewed, and the corpus is adjusted as necessary.

\textbf{Step2: Couple class model generation.} The couple class model must be constructed before the atomic class model in accordance with the principle of forward design through top-down model generation. Fine-tuned BERT for NER tags tokens, allowing us to identify system relationships within the simulation model corpus. This process ultimately enables the construction of the complete system composition by aggregating system relationships. The fine-tuned BERT model for single-sentence classification is capable of categorizing individual sentences. In Step 2, we employ it to determine whether the sentences in the model corpus describe the system’s connectivity relationships and add them to the connection corpus. Then, Tramsformer-based model inference is applied based on the connection corpus and the system composition to derive the connection relationships between subsystems. After defining the system composition and system connectivity relationships, the couple class model is constructed by populating the system composition into the keyword import of the couple class template and the system connection relations into the keyword connection.

\textbf{Step3: Atomic class model generation.} The name of the atomic class model corresponds to the name of the subsystem, and the keyword Port of the atomic class model is derived by analyzing the keyword Connection of the couple class model. Additionally, the valuetype of the port is reasoned through the port names and the model description document. Due to the complexity and variable format of the keyword State of the discrete class model, the keywords Value and State are coupled and generated by the fine-tuned Transformer-based model. The fine-tuning method for this model is described in detail in Section~\ref{subsec33}. The generation of the keyword State of the discrete class model should employ the Few-shot \cite{bib31} method to construct the prompts, as illustrated in Table.~\ref{mytab4}. Similar to the discrete class, the value and equation parts of the continuous class, which are coupled, are generated uniformly by Transformer-based models. After completing the construction of the state and equation parts, it is necessary to verify these components for any undefined functions. If there is an undefined function, the function class model of X language needs to be generated according to the function name and the code of the atomic class model. The function class supports the procedural programming paradigm so that the function class simulation code can be generated using GenAI for code scenarios.

\begin{algorithm}
\caption{Extract subsystem ports}\label{myalgo1}
\small
\begin{algorithmic}[1]
\State \textbf{Input:} keyword ``Connection'', subsystem name ``PartName''
\State \textbf{Output:} list of subsystem ports ``PortList''
\State Note: format of connect: $connect(part1.port1, part2.port2)$.
\State $PortList \gets \emptyset$
        \For{$connect$ \textbf{ in } $Connection$} 
            \If{$part1 == PartName$}
                \State $PortType \gets ReasonPortType(part1, port1)$
                \State $PortList.Add(''input '' + PortType + port1)$
            \EndIf
            \If{$part2 == PartName$}
                \State $PortType \gets ReasonPortType(part2, port2)$
                \State $PortList.Add(''output ''+ PortType + port2)$
            \EndIf
        \EndFor
\end{algorithmic}
\end{algorithm}

\begin{table}[h]
\caption{Method of constructing the prompts for generating the value and state parts}\label{mytab4}%
\renewcommand{\arraystretch}{1.2} 
\begin{tabular}{|m{0.8cm}|m{1.7cm}|m{5cm}|}
\hline
\textbf{From} & \textbf{Classification} & \textbf{Specific meaning} \\
\hline
user & BNF  & Backus-Naur Form (BNF) of X language    \\
\hline
user & state & Describe the parts that ``state'' contains and the text specification of ``state'', then give a few examples of ``state''  \\
\hline
system & state response & The Transformer-based model's understanding of the ``state'' specification   \\
\hline
user & introduction & ``Drawing on the textual descriptions of both the system model and the subsystem model, please develop the code for the keyword State of the discrete class subsystem model in accordance with the modeling specifications for the keyword State and the preceding code parts of the discrete class model. Note that only the code of the keyword State should be included in your output.''    \\
\hline
user & couple text   & Textual description of the couple class model to which the atomic class model to be generated belongs  \\
\hline
user & discrete text    & Textual description of the atomic class model to be generated   \\
\hline
user & generated code    & Generated code parts of the discrete class model, for example, keyword Name, keyword Parameter, keyword Port, etc.   \\
\hline
user & note    & Prompts added based on generic errors in Transformer-based model output results   \\
\hline
\end{tabular}
\end{table}

\subsection{NER-BERT Model Training}\label{subsec32}

This section details the training method for the NER task fine-tuned BERT model used in Step 2 of Section~\ref{subsec31}. The application of BERT \cite{bib32} is a pre-trained deep learning model that uses a bidirectional Transformer encoder to understand words in context and generate word vector representations with rich semantic information. One of the notable advantages of the BERT model lies in its high flexibility, allowing it to effectively adapt to various downstream tasks through a range of fine-tuning strategies \cite{bib33}.  NER-BERT presented in this section is employed to analyze the simulation model corpus in the early stages of model construction, identifying module containment relationships, extracting relevant information, and omitting irrelevant content. The extracted containment relationship data informs the construction of both the system model and its corresponding subsystems. 

\begin{figure}[!t]
\centerline{\includegraphics[width=\columnwidth]{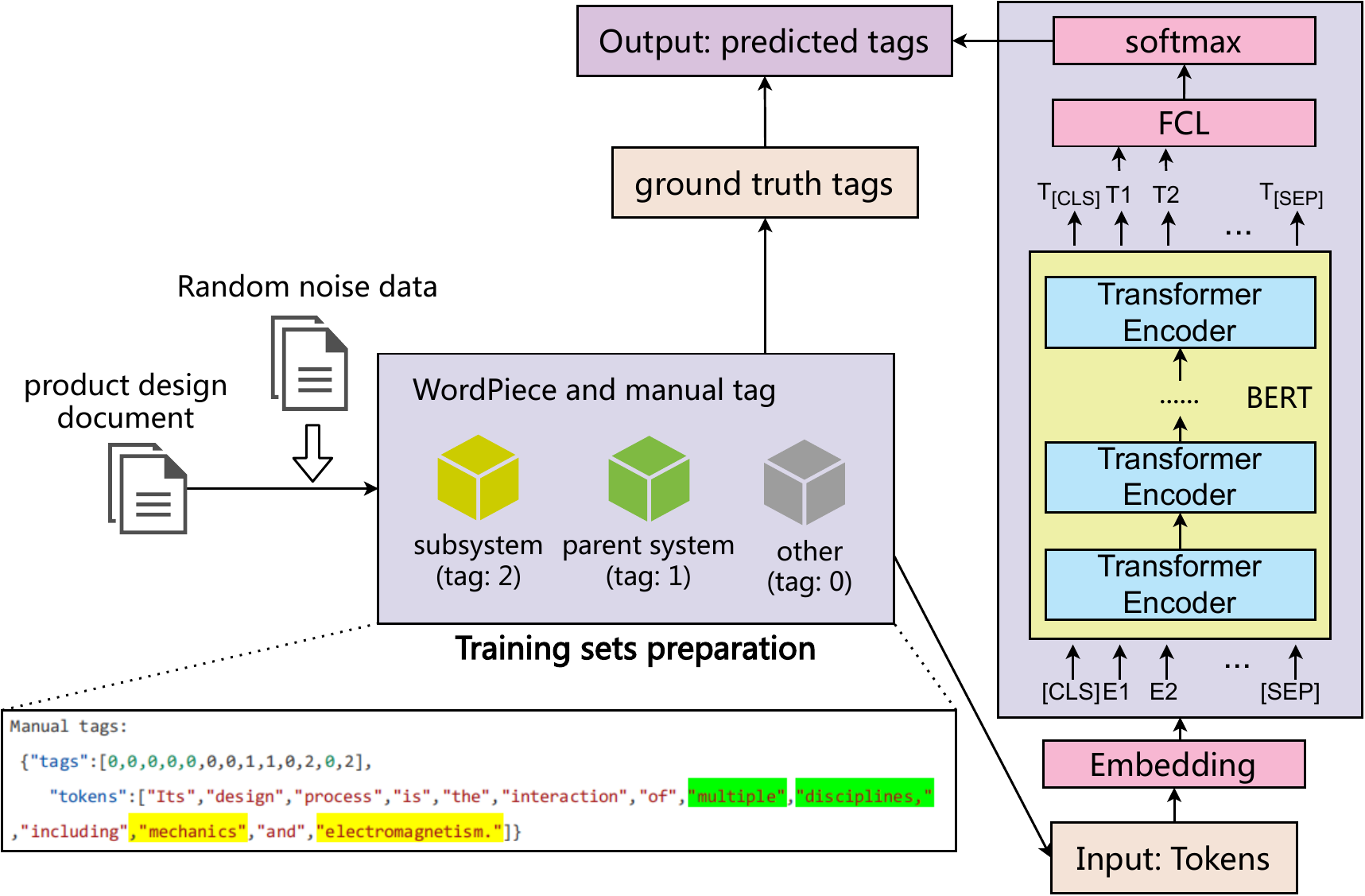}}
\caption{\raggedright The NER-BERT architecture comprises BERT, a fully connected layer, and a softmax layer. Multiple product design documents and random noise data are transformed into a training set through WordPiece and manual tagging. This dataset was subsequently used to fine-tune NER-BERT.}\label{myfig3}
\end{figure}

The training method of NER-BERT is illustrated in Fig.~\ref{myfig3}. When providing BERT with an input dataset, the constructed data source must meet the following conditions:
    1) It is the whole or part of the actual industrial model construction documentation;
    2) It contains a sufficient number of parent and subsystem containment relationships;
    3) It has substantial data without inclusion relationships as the data in question.

Upon completing the dataset preparation, fine-tuning training of the model can commence, enabling the BERT to perform the NER tasks outlined above. The input to BERT consists of the original word vectors for each word in the text, whereas the output comprises the vectors for each word or phrase after integrating the full semantic information of the text. As shown in the NER-BERT training section of Fig.~\ref{myfig3}, the fine-tuning process for the NER task connects a fully connected layer and a softmax layer to the original BERT output section. This addition enables the model to predict the tokens' tags based on these vectors. 

After training BERT for single-sentence classification scenarios in a similar manner, the system's composition and connection relations can be extracted using both NER-BERT and the fine-tuned BERT for single-sentence classification. These relations can then be used to construct X language couple class model.

\subsection{Transformer-based Models Training}\label{subsec33}

After constructing the couple class model, the next step is to generate the atomic class model, utilizing both the couple class model and the component-level model corpus. This section describes the training method for Transformer-based models, which is employed in Step 3 of Section~\ref{subsec31} for the code generation of the behavioral component of the atomic class model. The models trained in this section are primarily employed to generate the behavioral simulation codes within the atomic class models of X language, specifically the keyword State in discrete class models and the keyword Equation in continuous class models. 

\subsubsection{Data Set Preparation}\label{subsubsec231}

The dataset used to fine-tune Transformer-based was derived from X language model repository. However, this library faces two primary challenges: the limited number of models available for training large models and the significant homogenization of models. This homogenization arises because models with different structures and expressions have distinct application ranges. The scarcity of instances for some less frequently used models diminishes the quality of training. To address these issues, this paper employs generative modeling to create additional datasets, supplementing X language model library and providing a solution for the challenge of training corpus scarcity in practical applications.

\begin{figure}[!t]
\centerline{\includegraphics[width=\columnwidth]{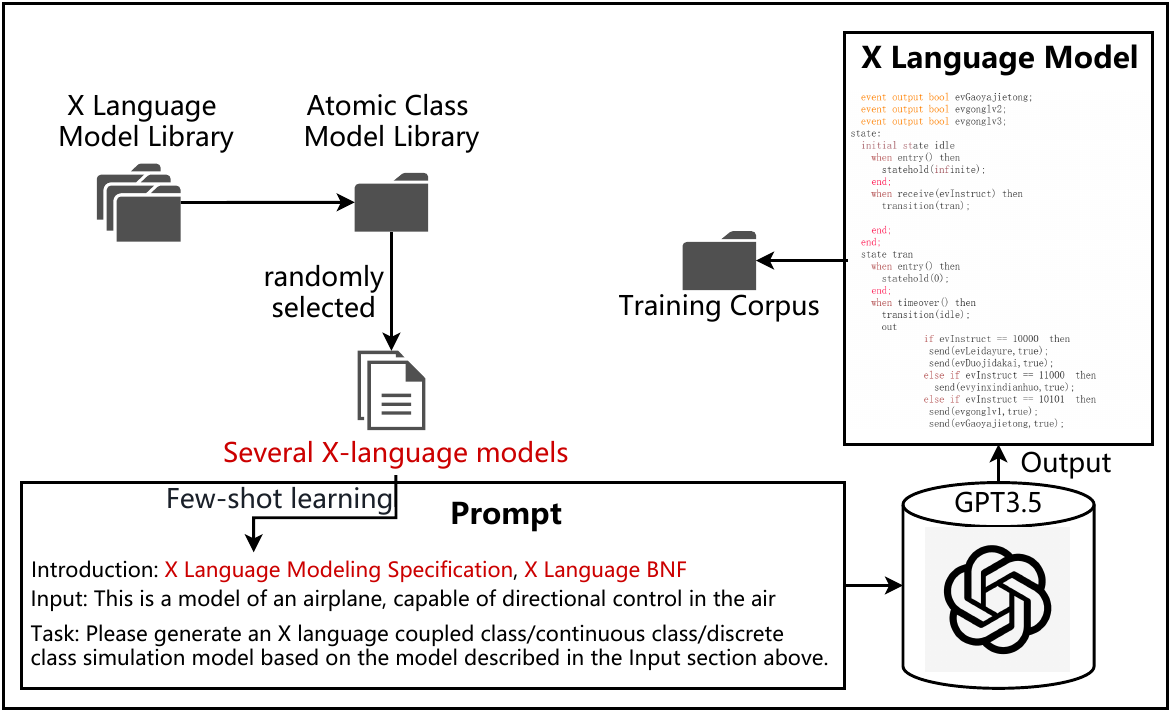}}
\caption{\raggedright This is an example of generating an X language simulation model. We extracted several X language models from X language model library and used them as samples to construct prompts using Few-shot learning. ChatGPT generates new X language models based on these prompts.}\label{myfig5}
\end{figure}

In this paper, we utilize GPT-3.5 for X language model generation. The prompts used to generate X language models consist of three components: Introduction, Input, and Task. The Introduction component encompasses prior knowledge of X language, including its modeling specifications and BNF, with several simulation models extracted from X language model library to facilitate Few-shot learning. The Input component represents the model description, which can be either a simple name or an extensive description, including, but not limited to, the model's function, parameters, and domain constraints. The Task component serves as the task description, i.e., “Generate a couple/continuous/discrete simulation model of X language based on the model described in the aforementioned Input component.” Fig.~\ref{myfig5} illustrates an example of generating an X language simulation model based on the prompts. The newly generated X language models, along with those in the original X language model library, constitute the training corpus. Subsequently, the contents of the corpus are converted into a dataset format suitable for training. The dataset format must include three components: Instruction, Input and Output. Taking the discrete class model as an example, the mask training method can be employed to achieve code completion for the keyword State of the discrete class model; that is, the other components of the discrete model are utilized to predict its state and value parts. Specifically, the Instruction, Input and Output components are presented in Table.~\ref{mytab2}. In addition to the Instruction texts listed in Table.~\ref{mytab2}, the Instruction for different samples adopts diverse and near-synonymous expressions to enhance the richness of the data and assist the training model in better understanding and generalizing the semantics of this specific instruction. The Output is extracted from the model using a regular expression method. 

\begin{table}[h]
\caption{Components of the dataset}\label{mytab2}%
\renewcommand{\arraystretch}{1.2} 
\begin{tabular}{|m{1.5cm}|m{6.4cm}|}
\hline
\textbf{Component} & \textbf{Content}  \\
\hline
Instruction    & ``This is a partially masked code for X language discrete class models. The parts represented by [MASK] may include value, state, and other keywords of discrete class models that have been concealed. Based on the available code, please speculate on the exact content in [MASK].''    \\
\hline
Input    & The value and state keywords of the discrete model are masked, with the masked parts replaced by ``[MASK]''.  \\
\hline
Output    & The code for the value and state parts of the discrete class model.   \\
\hline
\end{tabular}
\end{table}

\subsubsection{Transformer-based Model Training} \label{subsubsec232}

After the dataset's preparation, the model's training can commence. Specifically, let \( D \) represent the input simulation task description, 
\( C \) denotes the desired code result. Specifically, \( D \) represents the input in the dataset constructed above, whereas \( C \) represents the output in the same dataset. \( f(\cdot) \) signifies Transformer-based models. The output of Transformer-based models is:
\begin{align}
C' &=  f(D). \label{myeq1}
\end{align}
The optimization objective during training is to minimize the loss function \( \mathcal{L}(C', C) \), or, more precisely, maximize the probability of the conditional language model, as illustrated in \eqref{myeq2}. The goal is to maximize the log probability of predicting each word \(y_t\)in the target sequence \(y\), given the input sequence \(x\) and the previous words \(y<t\). The optimization is performed by adjusting the model parameters \(\Phi\) to improve the model’s prediction accuracy for the next word in the sequence. This section will focus on the fine-tuning of the model to achieve this objective. 
\begin{align}
\max _{\Phi} \sum_{(x, y) \in Z} \sum_{t=1}^{|y|} \log \left(P_{\Phi}\left(y_t \mid x, y<t\right)\right) \label{myeq2}
\end{align}

In this section, we choose  CodeQwen1.5-7B \cite{bib27} as the pre-trained model for fine-tuning. CodeQwen1.5 is a specialized code LLM built on the Qwen1.5 language model, and CodeQwen exhibits superior performance in long-term context understanding and generation compared to other open-source models of similar size. We utilizes the open-source project LLaMA Factory \cite{bib26} as the fine-tuning tool. LLaMA Factory is an LLM training and fine-tuning platform that offers greater flexibility and ease of use compared to other fine-tuning methods or tools. LLaMA Factory supports multiple model architectures and simplifies the fine-tuning process. We choose the Low-Rank Adaptation (LoRA) \cite{bib28} method to fine-tune the model. The LoRA method reduces the computational and memory requirements for training by decomposing the parameter matrices in a large model into two or more low-rank matrices and then updating only a portion of them. The principle of LoRA is shown in \eqref{myeq3}:
\begin{align}
\begin{split}
h=W_0 x+\Delta W x=W_0 x+B A x, \quad \\B \in \mathbf{R}^{d \times r}, A \in \mathbb{R}^{r \times d} \quad \text { and } r \ll \min (d, k). \label{myeq3}
\end{split}
\end{align}

\subsection{Evaluation Method for Simulation Models Generated by Transformer-based Models} \label{subsec34}

Mainstream code evaluation methods typically evaluate code based on metrics such as syntactic and semantic consistency (e.g., CodeBLEU \cite{bib43}) and functional correctness (e.g., Pass@k \cite{bib44}). However, simulation modeling of complex products introduces unique requirements and challenges for evaluation. In comparison to other code types, simulation model codes are characterized by their modular, hierarchical, and parametric structure. These characteristics cause the evaluation results of the above methods for evaluating simulation model codes to deviate from their actual quality. In contrast, the manual evaluation of simulation models, although capable of assigning reasonable scores based on the characteristics of the simulation model, ultimately depends on the scorer’s judgment in the absence of precise evaluation criteria. Therefore, this section introduces code evaluation metrics specific to X language simulation models tailored to their unique characteristics. These additions enhance the relevance of X language simulation model evaluations and provide specific metrics that guide manual evaluation. The evaluation metrics in this paper primarily evaluate X language simulation model based on Degree of Error and Model Consistency \cite{bib58}. These two metrics are described in detail below.

\textbf{Degree of error:} In product design scenarios, Simulation models with a low degree of error tend to require less effort during subsequent modifications. Thus, in addition to evaluating the accuracy of the generated model, the Degree of Error index of any erroneous model also holds a reference value. The Degree of Error differs from the n-gram \cite{bib45} approach. It evaluates the actual impact of errors on the model and assesses the ease of implementing modifications, and it often requires manual evaluation combined with a code checker and compiler.

\textbf{Model consistency:} Model consistency refers to the requirement that the representation of the same element in both the upper and lower models must remain consistent. The primary metrics of model consistency examined in this paper include:
    1) Name consistency between the attribute part of the system model and the header part of the subsystem
    2) Port consistency between the connection part of the system model and the definition part of the subsystem
    3) Consistency between the functions invoked in the atomic class model and the definitions provided in the function class model

Based on the above two metrics, the simulation model evaluation method proposed in this paper is outlined as follows:

A complete set of simulation models includes a top-level model and its subsystems, with each subsystem potentially containing additional subsystems. The parent model is always a couple class model, whereas sub-models may be an atomic class model. The score of the parent model is derived from the simulation correctness of itself and its sub-models, specifically:
\begin{align}
Score =   A_{parent} \cdot  \sum_{i=1}^{n} C_{subsystem,i} \cdot A_{subsystem,i}. \label{myeq5}
\end{align}
In \eqref{myeq5}, \(A_{parent}\) signifies the simulation correctness of the parent model (couple class), \(C_{subsystem,i}\) denotes the weight of the i-th subsystem model (couple class or atomic class),  and \(A_{subsystem,i}\) indicates the simulation correctness of the i-th subsystem model. For a set of simulation models undergoing evaluation, the final score is denoted as \(Score_{top}\), representing the score of the top-level model.

In this paper, the simulation correctness \( A_{i}\) is calculated as \eqref{myeq6}. X language simulation model is modular, allowing us to connect the ports of the model being tested to the correctly defined subsystem simulation models for test simulation. If the outputs of the model being tested are correct during the simulation, the model is considered fully correct.
\begin{align}
\mathrm{A}_{\mathrm{i}}=\left\{
\begin{array}{c c}
1, & \text{Fully correct} \\
\varepsilon \cdot P_i, & \text{Incorrect model simulation outputs}
\end{array}
\right. \label{myeq6}
\end{align}

In \eqref{myeq6}, \(\varepsilon\) represents penalty coefficient, a constant less than 1. \(P_i\) denotes the correctness similarity of the model. \(P_i\) is less than 1, with values closer to 1 indicating greater proximity to the correct model. The correctness similarity \( P_{c i}\) of the couple class model is calculated as \eqref{myeq7}.
\begin{align}
P_{ci}\!=  \!k_h \!\cdot \!P_{header,i} + k_a \!\cdot\! P_{attribute,i}  + k_c\! \cdot \!P_{connection,i}  \label{myeq7}
\end{align}
\(k_h,k_a\) and \(k_c\) denote the weights assigned to the correctness similarity for the attribute part and the connection part, respectively.\(P_{header,i}, P_{attribute,i}\) and \(P_{connection, i}\) denote the correctness similarity for the header part, the attribute part and connection part, respectively. \(P_{attribute,i}\) and \(P_{connection,i}\) are calculated by comparing the product design document with the attribute part and the connection part of the couple class model:
\begin{align}
P_{attribute,i}=F 1_{part,i} \cdot P_{port,i}, \label{myeq8}
\end{align}
\begin{align}
P_{connection,i}=F 1_{connection,i} \label{myeq9},
\end{align}
\(P_{port,i}\) denotes the correctness similarity for the keyword Port in the attribute part. \(F 1\) indicates F1 score in machine learning of keywords Part and Connection. The correctness similarity \( P_i\) of the atomic class model is calculated as \eqref{myeq11}.
\begin{align}
\begin{split}
P_i= & k_h \cdot P_{header,i}+k_d \cdot P_{definition,i}\\ &+ I(d)\cdot k_s \cdot P_{state,i} + I(c)\cdot k_e \cdot P_{equation,i}  \label{myeq11}
\end{split}
\end{align}

\(k_h\), \(k_d\), \(k_s\), and \(k_e\) denote the weights assigned to the correctness similarity for the header, definition, state (for discrete class) and equation parts (for continuous class), respectively. \(P_{header,i}\), \( P_{definition,i}\), \( P_{state,i}\), and \(P_{equation,i})\) denote the correctness similarity for the header part, definition part, state part (for discrete class), and equation part (for continuous class), respectively. \(I(d)\) and \(I(c)\)  represent indicator functions for the conditions ``model is a discrete class model'' and ``model is a continuous class model'' respectively, where the value is 1 when the condition is satisfied and 0 when it is not. \(P_{header,i}\), \( P_{port,i}\), \( P_{definition,i}\), \( P_{state,i}\), and \( P_{equation,i}\) are calculated as \eqref{myeq12} and \eqref{myeq14}.
\begin{align}
\begin{split}
\mathrm{P}_{\mathrm{*,i}}=&\left\{
\begin{array}{c c}
1, & \text{consistency} \\[0.5em]
\varepsilon_* \cdot \dfrac{CE_{*,i} }{CE_{*,i}+ IE_{*,i}}, & \text{inconsistency}
\end{array}
,\right. \\ &*=header, port, definition \label{myeq12}
\end{split}
\end{align}
\begin{align}
P_{*,i}= {\alpha_i}^m \cdot {\beta_i}^n, \hspace{0.2cm} *=state, equation  \label{myeq14}
\end{align}
\(\varepsilon_h\) and \(\varepsilon_d\) represent penalty coefficient. \(CE\) and \(IE\) denote elements that are consistent and inconsistent with other models, respectively. \({\alpha_i}\) and \({\beta_i}\) denote the attenuation of ``correctness similarity'' in the i-th model due to syntax errors and simulation logic errors, respectively, a constant less than 1. \(m\) and \(n\) denote the numbers of syntax errors and simulation logic errors in the model, respectively. Syntax detectors and compilers can typically detect syntax errors, whereas simulation logic errors must be identified through manual inspection. Consequently, from the perspective of modification effort, simulation logic errors are often more severe, necessitating that \({\beta_i}\) be smaller than \({\alpha_i}\). Considering that the Degree of Error in the state section is related to its length, the formulas for \({\alpha_i}\) and \({\beta_i}\) are provided to ensure that \(P_{state,i}\) is not influenced by the length of the keyword State:
\begin{align}
\log \alpha_i=\frac{\operatorname{len}_{\mathrm{c}}}{\operatorname{len}_{\mathrm{i}}} \cdot \log \alpha_c, \hspace{0.2cm} \log \beta_i=\frac{\operatorname{len}_{\mathrm{c}}}{\operatorname{len}_{\mathrm{i}}} \cdot \log \beta_c .  \label{myeq15}
\end{align}

\({\operatorname{len}_{\mathrm{c}}}\) and \({\operatorname{len}_{\mathrm{i}}}\) denote the standard keyword State length and the keyword State length of the i-th model, respectively. \(\alpha_c\) and \(\beta_c\) represent the standard attenuation coefficients. For the above evaluation metrics, \(P_{attribute,i}\), \( P_{header,i} \), and \(P_{definition,i}\) focus on the consistency of the model, whereas \(P_{state,i}\) and \(P_{equation,i}\) emphasize the Degree of Error in keyword State (for discrete class) and equation part (for continuous class) of the model.

The penalty coefficients in the above equation are determined based on the actual importance of each component of the model. The weights assigned to each model and to each component within the model are calculated using EWM. EWM is a multi-criteria decision-making approach that leverages the concept of information entropy to determine the weight of each criterion in an evaluation. This method assigns weights based on the variability of the data associated with each criterion; criteria with higher variability are assigned greater weights as they contribute more informative value to the evaluation \cite{bib59}.

This set of evaluation metrics considers the modularity of the simulation model and the characteristics of the design and development process. It will be employed in Section~\ref{sec4} to evaluate the quality of the generated simulation model.

\section{Simulation Model Generation for Aircraft Electrical System} \label{sec4}

This section generates and evaluates a set of simulation models for the aircraft electrical system based on the simulation model generation method outlined in Section~\ref{sec3}. Section~\ref{subsec41} details the experimental preparation required for generating the aircraft electrical system. Section~\ref{subsec42} demonstrates the efficacy of the fine-tuned NER-BERT model in extracting relevant information from the aircraft electrical system documentation. Section~\ref{subsec43} generates multiple sets of simulation models using different Transformer-based models. Section~\ref{subsec44}  evaluates the results against the model evaluation metrics presented in Section~\ref{subsec34}.

\subsection{Preparation of Experiments}\label{subsec41}

The preparation for the experiment will be presented in terms of case introduction, data preparation, and experimental equipment.

\begin{figure}[!t]
\centerline{\includegraphics[width=\columnwidth]{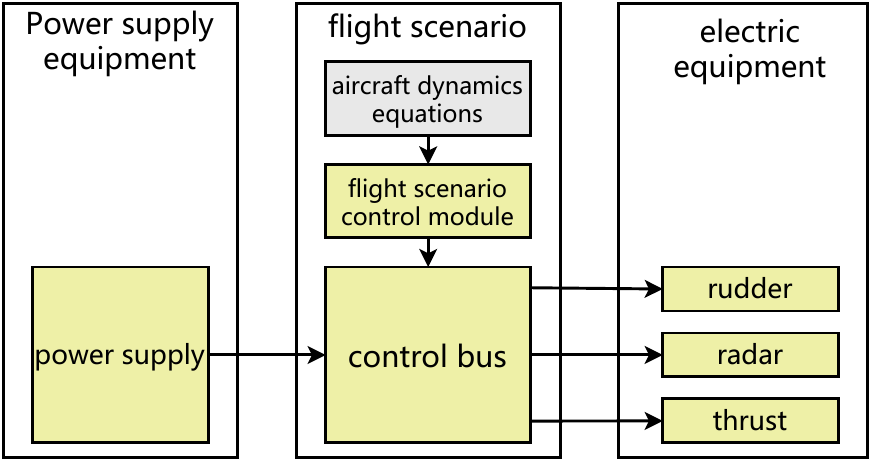}}
\caption{\raggedright The subsystems of the model are classified into three categories: power supply equipment, flight scenario control equipment, and power-using equipment. The flight scenario control equipment calculates the flight process parameters based on the aircraft dynamics equations and controls the power equipment via the control bus module.}\label{myfig18}
\end{figure}

\textbf{Case introduction:} The aircraft electrical system examines the variations in electrical current, voltage, and power utilized during the aircraft's actual flight. The aircraft electrical system comprises six components: power supply, flight scenario control module, control bus, radar, rudder, and thrust module. The relationship between the subsystems is illustrated in Fig.~\ref{myfig18}. The functions of these six subsystems are presented in Table.~\ref{mytab3}.

\begin{table}[h]
\caption{The functions of subsystems}\label{mytab3}%
\renewcommand{\arraystretch}{1.2} 
\begin{tabular}{|m{1.8cm}|m{6cm}|}
\hline
\textbf{Name} & \textbf{Function}  \\

\hline
power supply    & This subsystem provides a stable operating voltage of 28.5V to each module throughout the entire duration of operation.    \\
\hline
flight scenario control module    & This subsystem is responsible for designing flight trajectories that enable the aircraft to adhere to a predefined path.  \\
\hline
control bus    & This subsystem receives commands, parses them, and subsequently transmits the commands to the power equipment to adjust its power consumption.   \\
\hline
radar    & This subsystem detects the position of a target.    \\
\hline
rudder    & This subsystem adjusts power according to received commands.  \\
\hline
thrust    & This subsystem specializes in generating thrust to propel the aircraft forward.   \\
\hline
\end{tabular}
\end{table}

\textbf{Data preparation:} In this paper, we select relevant papers on the aircraft electrical system \cite{bib42,bib50} and the model-building instructions document compiled by our team for constructing the aircraft's electrical system as the data preparation for this experiment.

\textbf{Experimental equipment:} For the hardware configuration, the experimental computing platform includes an A100 80GB GPU, an i9-13900K processor, and 256GB of RAM. The software environment comprises Ubuntu 20.04 LTS as the operating system, PyTorch 2.4 as the deep learning framework, and CUDA 12.4 with cuDNN 9.1 for GPU acceleration.

\subsection{Couple Class Simulation Model Generation}\label{subsec42}

\begin{figure}[!t]
\centerline{\includegraphics[width=\columnwidth]{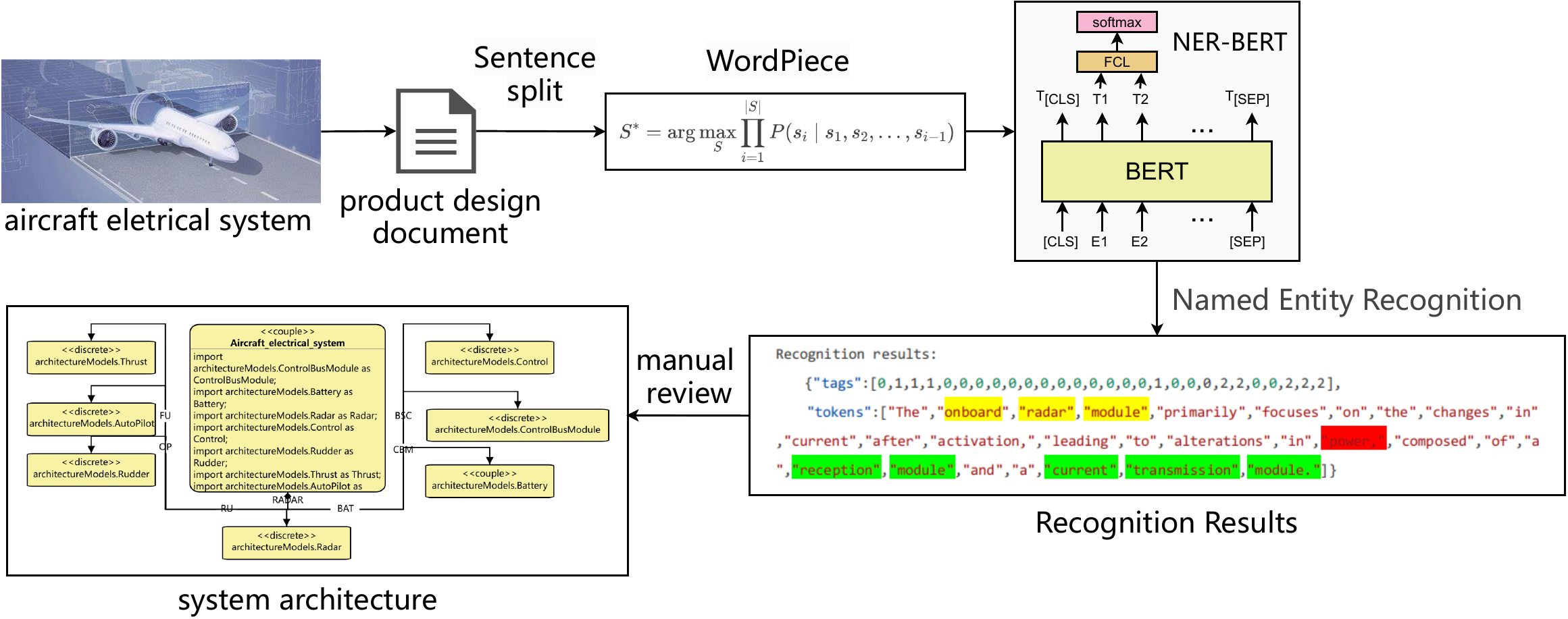}}
\caption{\raggedright The product design document is processed through sentence splitting and WordPiece tokenization, which are inputs to NER-BERT. NER-BERT classifies each token into one of the following categories: parent system (tagged as 1), subsystem (tagged as 2), or other (tagged as 0). Finally, the tags are manually reviewed, and the system composition of the model is derived based on these tags.}\label{myfig15}
\end{figure}

This section utilizes the NER-BERT model fine-tuned in Section~\ref{subsec32} to extract the system composition of the aircraft electrical system. The process is illustrated in Fig.~\ref{myfig15}. 

The tagging results of NER-BERT are compared with manual annotations, and if the tags for each token in a sentence are correct, NER-BERT is deemed to have correctly tagged the entire sentence. Ultimately, the accuracy of NER-BERT is 81.3\%.  Although this accuracy rate is relatively modest, we observe that the causes of incorrect token tags in NER-BERT can generally be categorized into two main groups:
    1) Errors in individual token tags within longer word sequences, i.e., incorrect boundaries of the tags.
    2) Misclassification of irrelevant tokens, which do not belong to any entity, as entity tags.
Specifically, the precision of entity tags for a given token is 95.06\%, whereas recall achieves 100\%. In addition to the aircraft electrical system, we also performed experiments on the electric vehicle system, railroad crossing system, and aircraft take-off system. The results of NER-BERT's entity tagging for these product design documents are presented in Table.~\ref{mytab421}. In the experiments, the recall rates for entity tags reached 100\%. This indicates that no system composition information is omitted, allowing engineers to adjust only the NER-BERT tagged content to derive the system composition, thus eliminating the need to read the entire product design document.

\begin{table}[htbp]
\centering
\caption{The results of NER-BERT's entity tagging} % 表格标题
\scriptsize
\begin{tabular}{ccccc} % c 表示居中对齐
\toprule % 顶部线
\textbf{Model Name} & \textbf{Sent Acc} & \textbf{Token Acc} & \textbf{Ent Prec} & \textbf{Ent Rec}  \\ 
\midrule % 表头下方的线
aircraft electrical system & 0.813 & 0.988 & 0.951 & \textbf{1}\\ 
electric vehicle system & 0.882 & 0.992 & 0.960 & \textbf{1}\\
railroad crossing system & 0.846 & 0.982 & 0.906 & \textbf{1}\\% 第一行数据
aircraft take-off system & 0.961 & 0.996 & 0.971 & \textbf{1}\\
\bottomrule % 底部线
\vspace{2mm}
\end{tabular}
\label{mytab421} % 表格标签，用于引用
\raggedright
\textbf{Notes:} \\
- Sent Acc: Sentence Accuracy - Token Acc: Token Accuracy \\
- Ent Prec: Precision of Entity Tags  - Ent Rec: Recall of Entity Tags
\end{table}

\subsection{Atomic Class Simulation Model Generation}\label{subsec43}

\begin{figure}[!t]
\centerline{\fbox{\includegraphics[width=\columnwidth]{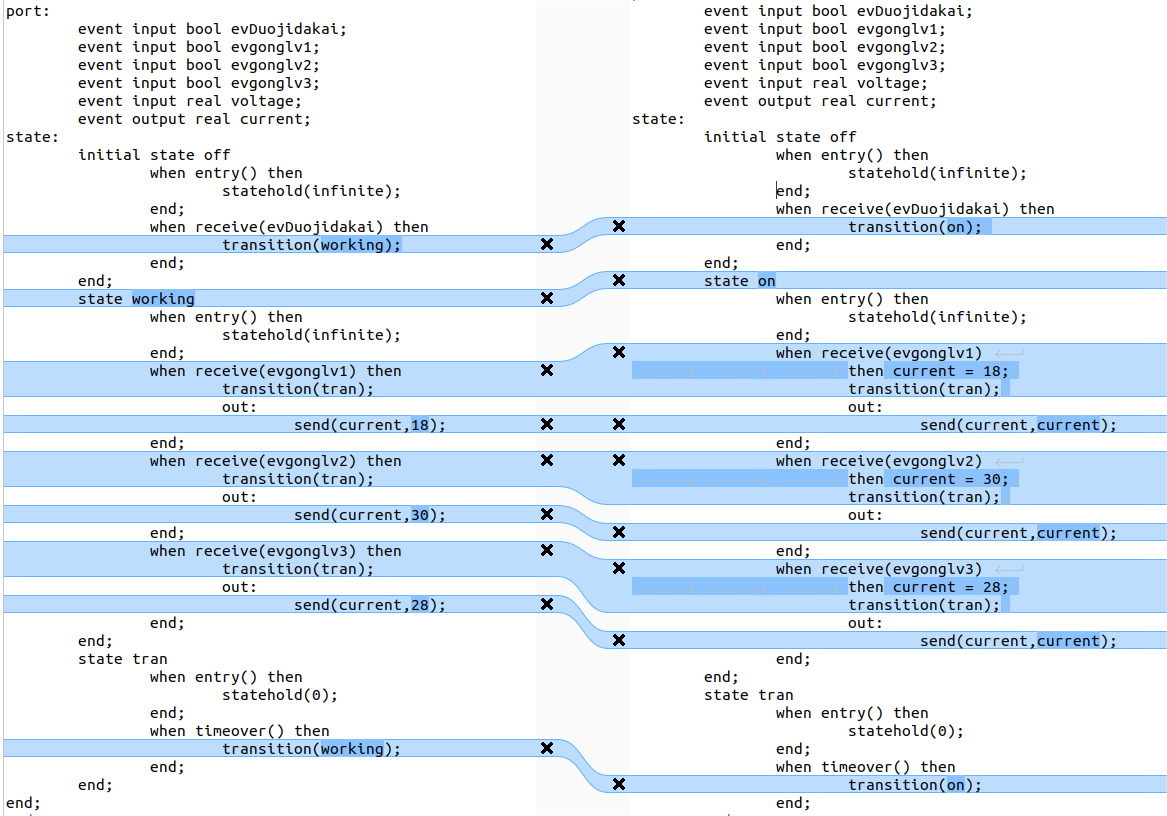}}}
\caption{\raggedright The left side of the image displays the generated model, whereas the right side shows the manually written model.}\label{myfig10}
\end{figure}

The atomic model generation method is presented in Step 3 of the Simulation Model Generation Method Based on Scalable Templates and Transformer-based Models in Section~\ref{subsec31}. Take the subsystem AutoPilot as an example. The AutoPilot subsystem performs the functions of the rudder module. Based on the connection part of the couple class model, AutoPilot can retrieve the relevant port connection relationships and populate the keyword Port of AutoPilot with this information. The prompts are formulated based on the structure outlined in Table.~\ref{mytab4} and serve as inputs to the Transformer-based models trained in Section~\ref{subsec33} to generate the behavioral code of the atomic class (state for discrete class, equation for continuous class) and its corresponding keyword Value. The comparison of generated model and manual written model is presented in Fig.~\ref{myfig10}.  Although there are differences in expression between the generated and manually written models, the generated model maintains the same semantics and adheres to X language syntax specifications.

\begin{figure}[!t]
\centerline{\includegraphics[width=\columnwidth]{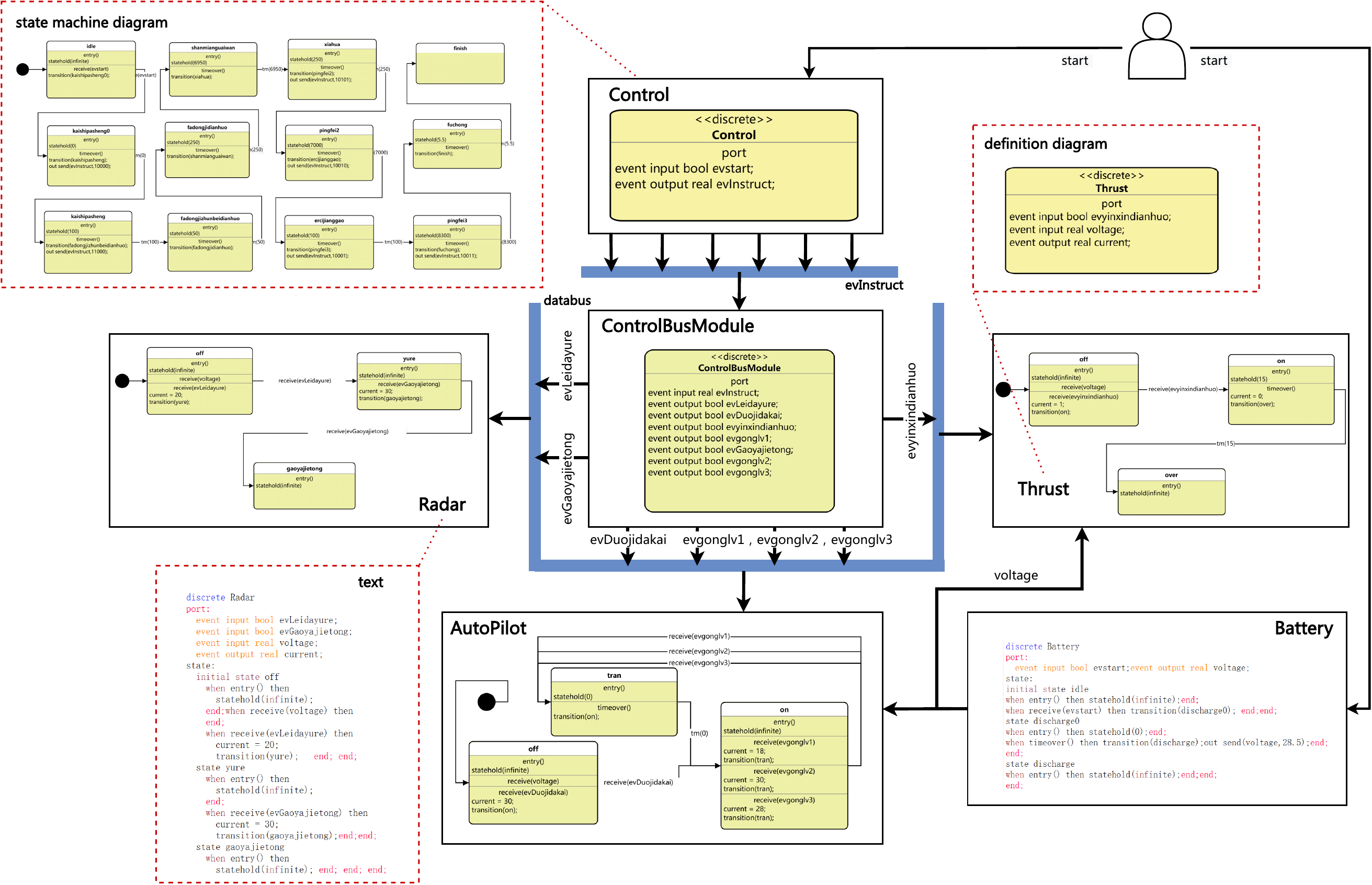}}
\caption{\centering The aircraft electrical system model}\label{myfig9}
\end{figure}

After completing the generation of all atomic models, both the generated couple class models and atomic models are imported into X language development tool, XLab. XLab can transform the generated text code into graphical models. In the aircraft electrical system model, each subsystem's graphical or textual composition, definitions, connectivity relationships, and internal logic are shown in Fig.~\ref{myfig9}. The mapping relationship between the final model and the various modules of the aircraft electrical system is outlined as follows: power supply - Battery, flight scenario control module - Control, control bus - BallisticSceneControl, radar - Radar, rudder - AutoPilot, thrust - Thrust. So far, we have successfully generated a complete simulation model of the aircraft electrical system using the Simulation Model Generation Method Based on Scalable Templates and Transformer-based Models proposed in this paper.

\subsection{Simulation Model Code Evaluation}\label{subsec44}

This section evaluates the impact of the aforementioned model generation methods on generating simulation models using various Transformer-based models based on the evaluation metrics outlined in Section~\ref{subsec34}.

We selected three open-source models for the code generation scenario (CodeQwen1.5-7b, CodeGemma-7b \cite{bib51}, and DeepSeek-Coder-6.7b \cite{bib52}) and evaluated the quality of the aircraft electrical system models generated with and without the use of the simulation model generation method proposed in this paper. Additionally, we selected two high-performance proprietary models (Claude-3.5-Sonnet, GPT-4o) and evaluated the aircraft electrical system models they generated directly. For the aircraft electrical system, set the values of \( \varepsilon\), \(\varepsilon_{header}\), \(\varepsilon_{port}\), and \( \varepsilon_{definition} \) in the evaluation metrics to 0.8, 0.6, 0.6, and 0.6, respectively.  Set the values of \( \alpha_c, \beta_c\), and \(len_c\) to 0.2, 0.1, and 1 state, respectively. We generated 20 sets of aircraft electrical system models using each model, both with and without the proposed method, and then manually evaluated them based on the evaluation metrics described above. 
We obtained several atomic class models and one couple class model using each method. We calculated the scores of atomic models as \ref{myeq7} and calculated their final scores as \eqref{myeq5}. We employed EWM to calculate the weights of the models and components, accentuating the gap in the final scores. The average scores of the 20 sets from different models and methods were compared, and the scores of atomic models and the final scores are presented in Table.~\ref{mytab431}. The meanings of Meth.a and Meth.b in Table.~\ref{mytab431} are as follows:
    Meth.a) The simulation model generation method proposed in this paper.
    Meth.b) Generate the simulation model directly using Transformer-based models. The prompts include X language BNF, X language model examples, a description of the model to be generated, and a description of the parent system model.

\begin{table*}[htbp]
\centering
\caption{Evaluation results of simulation models generated using different methods} % 表格标题
\small
\begin{tabular}{m{3.5cm}<{\centering} m{1cm}<{\centering} m{1cm}<{\centering} m{1cm}<{\centering} m{1cm}<{\centering} m{1cm}<{\centering} m{1cm}<{\centering} m{2cm}<{\centering} m{2cm}<{\centering} } % c 表示居中对齐
\toprule % 顶部线
\multirow{2}*{Model name} & \multicolumn{2}{c}{CodeGemma} & \multicolumn{2}{c}{CodeQwen} & \multicolumn{2}{c}{DeepSeek-Coder} & \multirow{2}*{Claude-3.5-Sonnic} &\multirow{2}*{GPT-4o} \\
\cline{2-3} \cline{4-5} \cline{6-7}
~ & Meth.a & Meth.b & Meth.a & Meth.b & Meth.a & Meth.b &  &  \\ 
\midrule % 表头下方的线
AutoPilot & 1 & 0.234 & 1 & 0.265 & 1 & 0.297 &0.402 & 0.379 \\% 第一行数据
BallisticSceneControl & 0.689 & 0 & 0.779 & 0.024 & 1 & 0.016 & 0.135 & 0.168\\ 
Battery & 1 & 0.474 & 1 & 0.435 & 1 & 0.457 & 0.485 & 0.51\\
ControlBusModule & 0.667 & 0 & 0.679 & 0 & 0.697 & 0 & 0 & 0.044\\
Thrust & 1 & 0.125 & 1 & 0.109 & 1 & 0.116 & 0.316 & 0.241\\
Oper & 0.823 & 0.263 & 0.828 & 0.255 & 0.881 & 0.260 & 0.257 & 0.266\\
Radar & 0.494 & 0.038 & 0.494 & 0.008 & 0.518 & 0.023 & 0.194 & 0.158\\
\midrule
Final score& \textbf{0.810} & 0.161 & \textbf{0.825} & 0.156 & \textbf{0.870} & 0.165 & 0.241 & 0.242 \\
\bottomrule % 底部线
\vspace{2mm}
\end{tabular}
\label{mytab431} % 表格标签，用于引用
\raggedright
\end{table*}

As shown in Table.~\ref{mytab431}, the simulation model generation method proposed in this paper improves the quality of the simulation models generated by mainstream Transformer-based models, such as CodeGemma, CodeQwen, and DeepSeek Coder. The quality of the generated simulation model surpasses that of directly using Claude 3.5, Sonnic, and GPT-4. This suggests a practical direction for generating simulation models focused on the physical properties of systems.

Additionally, we observed a phenomenon during our experiments: Transformer-based models perform poorly on the simple model Radar, which we believe is due to its a priori knowledge of radar. Instead of assisting in generating the simulation model, this prior knowledge hinders the process, which contradicts our expectations. This is most likely since the a priori knowledge related to radar in the Transformer-based models differs from the description of radar in the model design document. The Transformer-based models utilize their a priori knowledge when generating the simulation model, leading to a deviation between the final simulation model and the model design document. After renaming the Radar module, the quality of the simulation code for this module improved. This suggests that Transformer-based models' generalization ability may hinder the generation of high-quality simulation models when the simulation model is well documented. Determining how to help Transformer-based models better balance the application of prior knowledge with task requirements when generating content based on existing knowledge will be one of the future research directions for the application of GenAI in MBSE.

\section{Conclusion}\label{sec5}

This paper proposes a method for generating simulation models using scalable templates and Transformer-based models, offering a practical approach for the intelligent generation of simulation models of system physical properties in product systems. The main conclusions of this study are as follows:
    1) The token recognition in product design document sentences by the NER-BERT model trained in this study effectively filters redundant information and provides a reliable reference for designing the system composition.
    2) The simulation model evaluation method proposed in this study integrates the unique characteristics of simulation models and introduces evaluation metrics beyond code accuracy. This approach enables a quantitative evaluation of simulation codes.
    3) The method for generating simulation models using scalable templates and Transformer-based models generates simulation models for system physical properties through modular code completion, overcoming the limitations of LLM in handling long text inputs. The quality of simulation models generated with this method is improved compared to using Transformer-based models directly.

Although the Transformer-based models and scalable template-based simulation model generation methods proposed in this paper significantly enhance the simulation model generation code, there remains room for further improvement. In the future, more efforts will focus on expanding the size and diversity of the Transformer-based models training dataset and exploring the application of GenAI across other stages of the MBSE lifecycle.

\bibliographystyle{IEEEtran}
\bibliography{IEEEabrv,reference}

\begin{thebibliography}{10}
\providecommand{\url}[1]{#1}
\csname url@samestyle\endcsname
\providecommand{\newblock}{\relax}
\providecommand{\bibinfo}[2]{#2}
\providecommand{\BIBentrySTDinterwordspacing}{\spaceskip=0pt\relax}
\providecommand{\BIBentryALTinterwordstretchfactor}{4}
\providecommand{\BIBentryALTinterwordspacing}{\spaceskip=\fontdimen2\font plus
\BIBentryALTinterwordstretchfactor\fontdimen3\font minus \fontdimen4\font\relax}
\providecommand{\BIBforeignlanguage}[2]{{%
\expandafter\ifx\csname l@#1\endcsname\relax
\typeout{** WARNING: IEEEtran.bst: No hyphenation pattern has been}%
\typeout{** loaded for the language `#1'. Using the pattern for}%
\typeout{** the default language instead.}%
\else
\language=\csname l@#1\endcsname
\fi
#2}}
\providecommand{\BIBdecl}{\relax}
\BIBdecl

\bibitem{bib35}
S.~Friedenthal, R.~Griego, and M.~Sampson, ``Incose model based systems engineering (mbse) initiative,'' in \emph{INCOSE 2007 symposium}, vol.~11.\hskip 1em plus 0.5em minus 0.4em\relax sn, 2007.

\bibitem{bib10}
S.~Friedenthal, A.~Moore, and R.~Steiner, ``Omg systems modeling language (omg sysml) tutorial,'' in \emph{INCOSE Intl. Symp}, vol.~9.\hskip 1em plus 0.5em minus 0.4em\relax Citeseer, 2006, pp. 65--67.

\bibitem{bib11}
B.~Douglass, ``The {Harmony Process. I-Logix white paper, I-Logix},'' \emph{Inc.: Burlington, MA, USA}, 2005.

\bibitem{bib12}
J.~A. Estefan \emph{et~al.}, ``Survey of model-based systems engineering (mbse) methodologies,'' \emph{Incose MBSE Focus Group}, vol.~25, no.~8, pp. 1--12, 2007.

\bibitem{bib55}
X.~Li and J.~Liu, ``A method of sys ml-based visual transformation of system design-simulation models,'' \emph{Journal of Computer-Aided Design \& Computer Graphics}, vol.~28, no.~11, pp. 1973--1981, 2016.

\bibitem{bib54}
L.~Zhao, J.~Ye, H.~Qi, G.~Wei, and Z.~Yong, ``Simulation of civil aircraft takeoff scenario based on mbse,'' \emph{Journal of System Simulation}, vol.~33, no.~10, pp. 2499--2510, 2021.

\bibitem{bib46}
J.~Gray and B.~Rumpe, ``Reflections on the standardization of sysml 2,'' pp. 287--289, 2021.

\bibitem{bib47}
J.~Lu, G.~Wang, J.~Ma, D.~Kiritsis, H.~Zhang, and M.~T{\"o}rngren, ``General modeling language to support model-based systems engineering formalisms (part 1),'' in \emph{INCOSE international symposium}, vol.~30, no.~1.\hskip 1em plus 0.5em minus 0.4em\relax Wiley Online Library, 2020, pp. 323--338.

\bibitem{bib29}
B.~P. Zeigler, H.~Praehofer, and T.~G. Kim, \emph{Theory of modeling and simulation}.\hskip 1em plus 0.5em minus 0.4em\relax San Diego: Academic press, 2000.

\bibitem{bib17}
L.~Zhang, F.~Ye, Y.~Laili, K.~Xie, P.~Gu, X.~Wang, C.~Zhao, X.~Zhang, and M.~Chen, ``X language: an integrated intelligent modeling and simulation language for complex products,'' in \emph{2021 Annual Modeling and Simulation Conference (ANNSIM)}.\hskip 1em plus 0.5em minus 0.4em\relax IEEE, 2021, pp. 1--11.

\bibitem{bib18}
L.~Zhang, F.~Ye, K.~Xie, P.~Gu, X.~Wang, Y.~Laili, C.~Zhao, X.~Zhang, M.~Chen, T.~Lin \emph{et~al.}, ``An integrated intelligent modeling and simulation language for model-based systems engineering,'' \emph{Journal of Industrial Information Integration}, vol.~28, p. 100347, 2022.

\bibitem{bib2}
T.~B. Brown, ``Language models are few-shot learners,'' \emph{arXiv preprint arXiv:2005.14165}, 2020.

\bibitem{bib3}
H.~Touvron, L.~Martin, K.~Stone, P.~Albert, A.~Almahairi, Y.~Babaei, N.~Bashlykov, S.~Batra, P.~Bhargava, S.~Bhosale \emph{et~al.}, ``Llama 2: Open foundation and fine-tuned chat models,'' \emph{arXiv preprint arXiv:2307.09288}, 2023.

\bibitem{bib21}
J.~Achiam, S.~Adler, S.~Agarwal, L.~Ahmad, I.~Akkaya, F.~L. Aleman, D.~Almeida, J.~Altenschmidt, S.~Altman, S.~Anadkat \emph{et~al.}, ``Gpt-4 technical report,'' \emph{arXiv preprint arXiv:2303.08774}, 2023.

\bibitem{bib22}
B.~Roziere, J.~Gehring, F.~Gloeckle, S.~Sootla, I.~Gat, X.~E. Tan, Y.~Adi, J.~Liu, R.~Sauvestre, T.~Remez \emph{et~al.}, ``Code llama: Open foundation models for code,'' \emph{arXiv preprint arXiv:2308.12950}, 2023.

\bibitem{bib23}
Z.~Luo, C.~Xu, P.~Zhao, Q.~Sun, X.~Geng, W.~Hu, C.~Tao, J.~Ma, Q.~Lin, and D.~Jiang, ``Wizardcoder: Empowering code large language models with evol-instruct,'' \emph{arXiv preprint arXiv:2306.08568}, 2023.

\bibitem{bib24}
J.~Bai, S.~Bai, Y.~Chu, Z.~Cui, K.~Dang, X.~Deng, Y.~Fan, W.~Ge, Y.~Han, F.~Huang \emph{et~al.}, ``Qwen technical report,'' \emph{arXiv preprint arXiv:2309.16609}, 2023.

\bibitem{bib25}
Q.~Zheng, X.~Xia, X.~Zou, Y.~Dong, S.~Wang, Y.~Xue, Z.~Wang, L.~Shen, A.~Wang, Y.~Li \emph{et~al.}, ``Codegeex: A pre-trained model for code generation with multilingual evaluations on humaneval-x,'' \emph{arXiv preprint arXiv:2303.17568}, 2023.

\bibitem{bib30}
L.~Ren, H.~Wang, Y.~Tang, and C.~Yang, ``Aigc for industrial time series: From deep generative models to large generative models,'' \emph{arXiv preprint arXiv:2407.11480}, 2024.

\bibitem{bib57}
R.~Zhang, H.~Du, D.~Niyato, J.~Kang, Z.~Xiong, A.~Jamalipour, P.~Zhang, and D.~I. Kim, ``Generative ai for space-air-ground integrated networks,'' \emph{IEEE Wireless Communications}, 2024.

\bibitem{bib7}
J.~C{\'a}mara, J.~Troya, L.~Burgue{\~n}o, and A.~Vallecillo, ``On the assessment of generative ai in modeling tasks: an experience report with chatgpt and uml,'' \emph{Software and Systems Modeling}, vol.~22, no.~3, pp. 781--793, 2023.

\bibitem{bib9}
A.~Tikayat~Ray, B.~F. Cole, O.~J. Pinon~Fischer, A.~P. Bhat, R.~T. White, and D.~N. Mavris, ``Agile methodology for the standardization of engineering requirements using large language models,'' \emph{Systems}, vol.~11, no.~7, p. 352, 2023.

\bibitem{bib40}
E.~Bader, D.~Vereno, and C.~Neureiter, ``Facilitating user-centric model-based systems engineering using generative ai.'' in \emph{MODELSWARD}, 2024, pp. 371--377.

\bibitem{bib48}
T.-G. Kim and S.~Y. Lim, ``Hybrid modeling and simulation methodology based on devs formalism,'' in \emph{SCSC'2001}.\hskip 1em plus 0.5em minus 0.4em\relax ACM, 2001.

\bibitem{bib49}
G.~Wainer, ``Cd++: a toolkit to develop devs models,'' \emph{Software: Practice and Experience}, vol.~32, no.~13, pp. 1261--1306, 2002.

\bibitem{bib19}
Y.~Zhang, P.~Gu, Z.~Chen, and L.~Zhang, ``A method and implementation of automatic requirement tracking and verification for complex products based on x language,'' in \emph{China Intelligent Networked Things Conference}.\hskip 1em plus 0.5em minus 0.4em\relax Springer, 2022, pp. 443--455.

\bibitem{bib20}
P.~Gu, L.~Zhang, Z.~Chen, and J.~Ye, ``Collaborative design and simulation integrated method of civil aircraft take-off scenarios based on x language,'' \emph{Journal of System Simulation}, vol.~34, no.~5, pp. 929--943, 2022.

\bibitem{bib56}
K.~Xie, L.~Zhang, Y.~Laili, and X.~Wang, ``Xdevs: A hybrid system modeling framework,'' \emph{International Journal of Modeling, Simulation, and Scientific Computing}, vol.~13, no.~02, p. 2243001, 2022.

\bibitem{bib41}
A.~Vaswani, ``Attention is all you need,'' \emph{Advances in Neural Information Processing Systems}, 2017.

\bibitem{bib31}
O.~Vinyals, C.~Blundell, T.~Lillicrap, D.~Wierstra \emph{et~al.}, ``Matching networks for one shot learning,'' \emph{Advances in neural information processing systems}, vol.~29, 2016.

\bibitem{bib32}
J.~Lee and K.~Toutanova, ``Pre-training of deep bidirectional transformers for language understanding,'' \emph{arXiv preprint arXiv:1810.04805}, vol.~3, no.~8, 2018.

\bibitem{bib33}
Y.~Liu, ``Fine-tune bert for extractive summarization,'' \emph{arXiv preprint arXiv:1903.10318}, 2019.

\bibitem{bib27}
\BIBentryALTinterwordspacing
Q.~Team, ``Code with codeqwen1.5,'' April 2024. [Online]. Available: \url{https://qwenlm.github.io/blog/codeqwen1.5/}
\BIBentrySTDinterwordspacing

\bibitem{bib26}
Y.~Zheng, R.~Zhang, J.~Zhang, Y.~Ye, and Z.~Luo, ``Llamafactory: Unified efficient fine-tuning of 100+ language models,'' \emph{arXiv preprint arXiv:2403.13372}, 2024.

\bibitem{bib28}
E.~J. Hu, Y.~Shen, P.~Wallis, Z.~Allen-Zhu, Y.~Li, S.~Wang, L.~Wang, and W.~Chen, ``Lora: Low-rank adaptation of large language models,'' \emph{arXiv preprint arXiv:2106.09685}, 2021.

\bibitem{bib43}
S.~Ren, D.~Guo, S.~Lu, L.~Zhou, S.~Liu, D.~Tang, N.~Sundaresan, M.~Zhou, A.~Blanco, and S.~Ma, ``Codebleu: a method for automatic evaluation of code synthesis,'' \emph{arXiv preprint arXiv:2009.10297}, 2020.

\bibitem{bib44}
C.~Raffel, N.~Shazeer, A.~Roberts, K.~Lee, S.~Narang, M.~Matena, Y.~Zhou, W.~Li, and P.~J. Liu, ``Exploring the limits of transfer learning with a unified text-to-text transformer,'' \emph{Journal of machine learning research}, vol.~21, no. 140, pp. 1--67, 2020.

\bibitem{bib58}
O.~Iordache, \emph{Modeling multi-level systems}.\hskip 1em plus 0.5em minus 0.4em\relax Springer Science \& Business Media, 2011.

\bibitem{bib45}
M.~Chen, J.~Tworek, H.~Jun, Q.~Yuan, H.~P. D.~O. Pinto, J.~Kaplan, H.~Edwards, Y.~Burda, N.~Joseph, G.~Brockman \emph{et~al.}, ``Evaluating large language models trained on code,'' \emph{arXiv preprint arXiv:2107.03374}, 2021.

\bibitem{bib59}
Y.~Zhu, D.~Tian, and F.~Yan, ``Effectiveness of entropy weight method in decision-making,'' \emph{Mathematical Problems in Engineering}, vol. 2020, no.~1, p. 3564835, 2020.

\bibitem{bib42}
P.~Gu, Y.~Zhang, Z.~Chen, C.~Zhao, K.~Xie, Z.~Wu, and L.~Zhang, ``X-rmtv: An integrated approach for requirement modeling, traceability management, and verification in mbse,'' \emph{Systems}, vol.~12, no.~10, p. 443, 2024.

\bibitem{bib50}
P.~Gu, Y.~Li, Z.~Wu, Z.~Chen, K.~Xie, and L.~Zhang, ``Research on integrated modeling and simulation method of missile electrical system based on x language,'' in \emph{China Intelligent Networked Things Conference}.\hskip 1em plus 0.5em minus 0.4em\relax Springer, 2024, pp. 219--230.

\bibitem{bib51}
C.~Team, H.~Zhao, J.~Hui, J.~Howland, N.~Nguyen, S.~Zuo, A.~Hu, C.~A. Choquette-Choo, J.~Shen, J.~Kelley \emph{et~al.}, ``Codegemma: Open code models based on gemma,'' \emph{arXiv preprint arXiv:2406.11409}, 2024.

\bibitem{bib52}
D.~Guo, Q.~Zhu, D.~Yang, Z.~Xie, K.~Dong, W.~Zhang, G.~Chen, X.~Bi, Y.~Wu, Y.~Li \emph{et~al.}, ``Deepseek-coder: When the large language model meets programming--the rise of code intelligence,'' \emph{arXiv preprint arXiv:2401.14196}, 2024.

\end{thebibliography}

\end{document}